\documentclass{IEEEtran}
\usepackage{amsmath}  
\usepackage{amsfonts} 
\usepackage{textcomp}
\usepackage[utf8]{inputenc} 
\usepackage[T1]{fontenc}    
\usepackage{hyperref}       
\usepackage{url}            
\usepackage{booktabs}       
\usepackage{amsfonts}       
\usepackage{nicefrac}       
\usepackage{microtype}      
\usepackage{lipsum}
\usepackage[pdftex]{graphicx}
\usepackage{amssymb,amsmath, dsfont, mathtools, amsthm}
\usepackage{xcolor}
\usepackage{mathrsfs}
\usepackage{tensor}
\usepackage{bm}
\usepackage{cuted}

\newcommand{\RR}{\mathbb{R}}
\newcommand{\dd}{\mathrm{d}}
\newcommand{\ee}{\mathrm{e}}
\newcommand{\ii}{\mathrm{i}}

\newtheorem{definition}{Definition}[section]

\newtheorem*{remark}{Remark}
\usepackage{float}
\usepackage{tikz-cd}
\usepackage{amsthm}

\begin{document}

\title{\fontsize{16}{18}\selectfont
Electromagnetism in Curved Spacetimes: Coupling of the Doppler and Gravitational Redshifts}
\author{Cameron R. D. Bunney, Gabriele Gradoni, \emph{Member, IEEE}
\thanks{Cameron R. D. Bunney is with the School of Mathematical Sciences, University of Nottingham (e-mail: cameron.bunney@nottingham.ac.uk). }
\thanks{Gabriele Gradoni is with the School of Mathematical Sciences, University of Nottingham and George Green Institute for Electromagnetics Research, University of Nottingham (e-mail: gabriele.gradoni@nottingham.ac.uk).}}

\maketitle





\begin{abstract}
We present the basic prerequisites of electromagnetism in flat spacetime and provide the description of electromagnetism in terms of the Faraday tensor. 
We generalise electromagnetic theory to a general relativistic setting, introducing the Einstein field equations to describe the  propagation of electromagnetic radiation in curved space-time. 
We investigate gravitational redshift and derive formulae for the combined effect of gravitational redshift and the Doppler shift in curved spacetime.
\end{abstract}


\section{Introduction}
The general relativistic treatment of electromagnetism is typically developed by promoting Minkowski spacetime to a general, possibly curved, spacetime, which is more physically relevant since matter curves spacetime in the universe.

The current pedagogical works on relativistic electrodynamics come from two points of view: 
the first is for those with a background in electromagnetism and deals with introductory special relativity; 
the second is for those with a background in gravity, which studies curved spacetimes and general relativity, with examples involving electromagnetism. 
The aim of this article is to review these two areas as a joint theory for the applied electromagnetism audience. 

We begin by introducing index notation; the language of relativity relies on this to condense equations. 
Without this notation, the Einstein field equations would instead be written as $10$ separate partial differential equations, 
as opposed to a single equation. 
We then present the reformulation of electrodynamics in terms of the \textit{Faraday tensor}, an object key to all relativistic electrodynamics. After a small exposition of electromagnetism in flat spacetime, we begin with general relativity. Generalisations of the derivative to the covariant derivative and the process of minimal coupling are introduced, bringing Maxwell’s equations in terms of the Faraday tensor to general relativistic form. Curvature is introduced and motivated by comparison of Schwarz’s theorem for second partial derivatives to its covariant form, which doesn't hold in general. We present the energy-momentum tensor, an object at the heart of general relativity, coupling matter to electrodynamics. This tensor, when specialised to electromagnetism, includes the Poynting vector and the Maxwell stress tensor, 
among other physical quantities that are relevant in the Einstein equations. 
Due to properties of the electromagnetic energy-momentum tensor, we are able to simplify Einstein’s field equations to the Maxwell-Einstein field equations. 
We then discuss the Lorenz gauge and the canonical wave equation in this theory, yielding the de Rham wave equation. We provide a solution to this, as in \cite{Gravitation}, through the geometrical optics approximation. The discussion again moves to discuss electrodynamics in curved spacetimes. Key concepts such as the four-velocity and proper time are explained, leading to the key dynamical equation of motion. 
Changing from Euclidean space to a curved spacetime, the geodesic 
equation is found in an approximated form. 
We investigate the dipole radiation in \cite{Griffiths} as motivation to study the propagation of radiation in curved spacetimes, giving an analysis of the geodesics that light and therefore electromagnetic waves travels along in a general, spherically symmetric spacetime. 
We further the study of \textit{null ray} solutions to the calculation for non-radial null rays. This investigation allows us to delve into gravitational redshift, re-deriving a well-known approximation \cite{CarrollAstronomy}, the redshift of light sent between two stationary observers. We then derive results for the redshift of light experiencing both gravitational redshift and the Doppler shift, sent between two observers which may be in $1)$ radial motion, or $2)$ a combination of both radial and azimuthal motion. 
\section{Electromagnetism in Flat Spacetime}
We begin by briefly recalling Maxwell's equations with $c= 1/\sqrt{\varepsilon_0 \mu_0} = 1$ and with metric convention $(-,+,+,+)$ in flat spacetime, 
meaning that all line elements/metrics will be of the form $\dd s^2 = - F(\textsf{x}) \dd t^2 + \dd s^2_{space}$, where $F(\textsf{x})$ is some function of potentially all coordinates. We use $c=1$, natural units, to drop prefactors of $c$, which can be restored via dimensional analysis. In particular for electromagnetism in flat spacetime, we consider Minkowski space in coordinates with line element $\dd s^2 = -\dd t^2 + \dd x^2 + \dd y^2 + \dd z^2$.
This leads to
\begin{align*}
    \nabla \cdot \bm{E} &= \frac{\rho}{\varepsilon_0}, &
    \nabla \cdot \bm{B} &= 0, \\
    \nabla \times \bm{E} &= -\frac{\partial \bm{B}}{\partial t}, &
    \nabla \times \bm{B} &= \mu_0 \bm{J} + \frac{\partial \bm{E}}{\partial t}.
\end{align*}
\subsection{Index Notation}
By defining a so-called four-current, $J^\mu = \left(\rho, \bm{J}\right)^\mu$, introducing the derivative notation $\partial_\mu = (\partial_t, \bm{\nabla})_\mu$ and the Levi-Civita symbol, one  reformulates Maxwell's equations into a compact form. The Levi-Civita symbol, also known as the alternating tensor is defined to be
\begin{equation*}
    \varepsilon^{ijk} = 
\begin{cases} 
        1 & \text{if }ijk \text{ an even permutation of }123, \\
        -1 & \text{if }ijk \text{ an odd permutation of }123, \\
      0 & \text{otherwise.}
\end{cases}
\end{equation*}
An even permutation is where cyclic permutations of $(123)$ are performed, whereas odd permutations involve flipping two elements, then followed by cyclic permutation, $(132)$.
Index, or suffix notation, is a way of writing equations without writing out sums explicitly. The general rule is that if an index is in both the upper and lower positions, one goes ahead ans sums over it. Usually, Roman indices run from $1$ to $3$, and Greek run from $0$ to $3$. The cross product between two vectors can be written as
\begin{equation*}
    (\bm{a}\times \bm{b})^i = \varepsilon^{ijk}a_j b_k.
\end{equation*}This can be verified using the definition of the Levi-Civita symbol above. One may then write the differential operations of curl and divergence using this convention. These can be applied to rewrite out the Maxwell equations in index form
\begin{align*}
    \partial_i E^i &= \frac{J^0}{\varepsilon_0},&
    \partial_i B^i &= 0, \\
    \varepsilon^{ijk}\partial_jE_k &= - \partial_0B^i,&
    \varepsilon^{ijk}\partial_jB_k &= \mu_0 J^i + \partial_0E^i.
\end{align*}
\begin{remark}
Partial derivatives are denoted $\partial_i = \partial/\partial x^i$.
\end{remark}
\subsection{Faraday Tensor}
Since the divergence of the magnetic field is always zero, one may write this in terms of a vector potential,
\begin{equation}\label{eqn: B potential}
    \bm{B} = \bm{\nabla}\times\bm{A}.
\end{equation}
Similarly, one may write
\begin{equation}\label{eqn: E potential}
    \bm{E} = -\bm{\nabla}\phi-\frac{\partial}{\partial t}\bm{A}.
\end{equation}
Both fields are fully described by using only a vector and a scalar potential. One may consider combining both potentials to make a \textit{four-potential} to be a possible tool to describing electromagnetism in a relativistic notation:
\begin{align*}
    A^\mu = (\phi, \bm{A})^\mu, & \quad A_\mu = (-\phi, \bm{A})_\mu.
\end{align*}
Minkowski space, as before, has the line element (in coordinates $(t,x,y,z)$) $\dd s^2 = -\dd t^2 + \dd x^2+\dd y^2+\dd z^2$, which can be compactly written as $\dd s^2 = \eta_{\mu\nu}\dd x^\mu \dd x^\nu$ with $\eta_{\mu\nu}$ a diagonal matrix, $\eta_{\mu\nu} = \mathrm{diag}(-1,1,1,1)_{\mu\nu}$. One often calls $\eta_{\mu\nu}$ the \textit{metric} as $ds^2$ and $\eta_{\mu\nu}$ can be derived from each other. For general spacetimes, one writes $\dd s^2 = g_{\mu\nu}\dd x^\mu \dd x^\nu$ as a convention. Here, $g_{\mu\nu}$ when considered as a matrix is a symmetric matrix. Objects with two indices can be written as matrices and objects with one index can be written as row or column vectors. An object with indices is a tensor if it obeys a certain transformation law which can be found in any introductory general relativity textbook, including \cite{CarrollBook}. Whether a tensor has indices up, down, or both, determines the nomenclature surround it. For example an example $C^\mu$ is a $(1,0)$-tensor, whereas $C_\mu$ is a $(0,1)$-tensor and $C^\mu_\nu$ is a $(1,1)$-tensor, if these three objects obey the right transformation rules. By using the metric $g_{\mu\nu}$ or the inverse metric $g^{\mu\nu}$, there is a way of moving indices up and down. The inverse metric is defined as the $(2,0)$-tensor such that
\begin{equation*}
    g_{\mu\nu}g^{\nu\rho} = \delta^\rho_\mu,
\end{equation*}where $\delta^\rho_\mu$ is the Kronecker delta, serving as the index equivalent of the identity matrix. One can intuitively think of this in terms of matrices. Consider matrices $A$ and $B$, then $(AB)_{ij} = \sum_k A_{ik}B_{kj}$. The use of Einstein summation follows exactly what we would expect.

In relativity, keeping track of indices is important such as whether an index is up or down. The act of moving an index from down to up is called \textit{raising} and vice versa is called \textit{lowering}. Consider $A^\mu$ and $B_\mu$, we raise and lower as follows:
\begin{align*}
    A_\nu := g_{\mu\nu}A^\mu,& &
    B^\nu := g^{\mu\nu}B_\mu.
\end{align*}
The power of writing the potentials in one four-potential is in an object defined as
\begin{equation}\label{eq: Faraday}
    F^{\mu\nu} = \partial^\mu A^\nu - \partial^\nu A^\mu.
\end{equation}
This $(2,0)$-antisymmetric tensor is called the Faraday tensor and has components
\begin{equation*}
    F^{\mu\nu} = 
    \begin{pmatrix}
    0 & E^1 & E^2 & E^3 \\
    -E^1 & 0 & B^3 & -B^2 \\
    -E^2 & -B^3 & 0 & B^1 \\
    -E^3 & B^2 & -B^1 & 0
    \end{pmatrix}^{\mu\nu}
    = -F^{\nu \mu},
\end{equation*} and Maxwell's equations may be reduced down to two using the Faraday tensor,
\begin{align}
    \partial_\mu F^{\nu \mu} = \mu_0 J^\nu, \label{eqn:dFJ}\\
     \partial_{[\sigma} F_{\mu \nu]} = 0. \label{eqn:dF}
\end{align}
Here, the square brackets represent the \textit{anti-symmetrisation} of the tensor. On a $(0,n)$-tensor, this is defined as
\begin{align*}
    A_{\left[\alpha_1 \alpha_2 \dots \alpha_n\right]} =& \frac{1}{n!}(A_{\alpha_1\alpha_2\dots \alpha_n} + A_{\alpha_2 \dots \alpha_n \alpha_1} \\
    &+ \dots \text{All even permutations} \\
    &- A_{\alpha_2 \alpha_1 \dots \alpha_n} - A_{\alpha_1 \dots \alpha_n \alpha_2}\\
    &- \dots \text{All odd permutations}).
\end{align*}
The \textit{symmetrisation} can be similarly defined by writing $A_{(\alpha_1 \alpha_2 \dots \alpha_n)}$ and changing the signs on the odd permutations to "$+$" signs.
In particular, in the equation above, we have
\begin{align*}
    \partial_{\left[\sigma\right.}F_{\left.\mu\nu\right]} = \frac{1}{3!}(&\partial_\sigma F_{\mu\nu} + \partial_\mu F_{\nu\sigma} + \partial_\nu F_{\sigma\mu} \\
    -&\partial_\nu F_{\mu\sigma} - \partial_\mu F_{\sigma \nu} - \partial_\sigma F_{\nu\mu} ) .
\end{align*}
\begin{remark}
It is very common to use $\eta_{\mu \nu} = diag(1,-1,-1,-1)_{\mu\nu}$ instead as a metric. In this case this would have its own Faraday tensor. The relationship between the Faraday tensor used here and the other convention is 
\begin{equation*}
    F^{\mu \nu}_1 = - F^{\mu \nu}_3,
\end{equation*}
where the subscript $1$ represents the single minus sign in our convention and the $3$ represents the three minus signs in the other convention.
\end{remark}
\section{Electromagnetism in Curved Spacetimes}
The theory of electromagnetism in Minkowski space generalises to the case where the spacetime has curvature. The method by which the equations are coupled to gravity in the curved spacetime is (often) by replacing partial derivatives by \textit{covariant derivatives} $\nabla_\mu$ and the Minkowski metric $\eta_{\mu\nu}$ by a general metric $g_{\mu\nu}$ in a process called \textit{minimal coupling}.
When acting on vectors, the partial derivatives become covariant, defined as
\begin{equation*}
    \partial_\mu A^\nu \mapsto \nabla_\mu A^\nu = \partial_\mu A^\nu + \Gamma^\nu_{\mu\rho}A^\rho,
\end{equation*}
and when acting on covectors, one has
\begin{equation*}
    \partial_\mu B_\nu \mapsto \nabla_\mu B_\nu = \partial_\mu B_\nu - \Gamma^\rho_{\mu\nu}B_\rho.
\end{equation*}
Here, $\Gamma^\nu_{\mu\rho}$ are the \textit{Christoffel symbols} defined by
\begin{equation*}
    \Gamma^\nu_{\mu\rho} = \frac{1}{2}g^{\nu\alpha}\left(\partial_\mu g_{\rho\alpha} + \partial_\rho g_{\mu\alpha} - \partial_\alpha g_{\mu\rho} \right).
\end{equation*}
The covariant derivative arises since the standard partial derivative does not have the good transformation rules of a tensor when applied to anything other than functions in a curved spacetime. However, for scalars, which have no free indices, we define the covariant derivative to be
\begin{equation*}
\nabla_\mu \chi = \partial_\mu \chi.
\end{equation*}
Furthermore, one demands that the covariant derivative satisfies:
\begin{align*}
    i) &\textit{ Metric Compatibility: } \nabla_\mu g_{\nu\rho} = 0, \\
    ii) &\textit{ Symmetry: } \Gamma^\nu_{\mu \rho} = \Gamma^\nu_{\rho\mu}.
\end{align*}This is called the Levi-Civita connection and it is uniquely defined by the Christoffel symbols. One may apply minimal coupling first to the Faraday tensor as such
\begin{align*}
    F^{\mu\nu} &= g^{\mu\rho}g^{\nu \sigma}F_{\rho\sigma},\\
    &=g^{\mu\rho}g^{\nu \sigma} \left(\nabla_\rho A_\sigma - \nabla_\sigma A_\rho \right),\\
    &=g^{\mu\rho}g^{\nu \sigma} \left(\partial_\rho A_\sigma - \partial_\sigma A_\rho - \Gamma^\alpha_{\rho \sigma} + \Gamma^\alpha_{\sigma \rho} \right),\\
    &= g^{\mu\rho}g^{\nu \sigma} \left(\partial_\rho A_\sigma - \partial_\sigma A_\rho\right), \\
    &= \partial^\mu A^\nu - \partial^\nu A^\mu.
\end{align*}Here we once again retrieve the flat spacetime definition of the Faraday tensor as in \eqref{eq: Faraday}.

\begin{remark}
One may apply minimal coupling to the same gauge symmetry on the vector potential as before
\begin{equation*}
    A^\mu \mapsto A^\mu + \nabla^\mu \chi,\quad \chi \in C^2\left( M\right)
\end{equation*}
where $C^2\left( M\right)$ is the space of twice differentiable functions on our spacetime $M$, and due to the equivalence of partial and covariant derivatives on scalar functions, this is exactly the same transformation as before.
\end{remark}
By the process of minimal coupling, one also receives the curved spacetime generalisations of the Maxwell equations in \eqref{eqn:dFJ} and \eqref{eqn:dF}.
\subsection{Curvature and Einstein's Field Equations}
A curved spacetime is characterized by its curvature. 
To start, consider Schwarz's theorem (partial derivatives commute)  and use our substitution rule
\begin{equation*}
    \partial_\mu\partial_\nu \mapsto \nabla_\mu \nabla_\nu,
\end{equation*}with $\partial_\mu\partial_\nu = \partial_\nu\partial_\mu$. 
If we act on some covector field $B_\alpha$, it can be shown that 
\begin{align*}
    \nabla_\mu\nabla_\nu B_\alpha - \nabla_\nu\nabla_\mu B_\alpha = (\partial_\nu \Gamma^\alpha_{\mu\rho} - \partial_\mu \Gamma^\alpha_{\nu \rho} + \Gamma^\beta_{\mu\rho}&\Gamma^\alpha_{\nu\beta}\\- \Gamma^\beta_{\nu\rho}&\Gamma^\alpha_{\mu\alpha} )B_\alpha.
\end{align*}
which in general is non zero. In fact, we define a new tensor, the Riemann curvature tensor by
\begin{equation}\label{RiemannLower}
    \nabla_\mu\nabla_\nu B_\rho - \nabla_\nu\nabla_\mu B_\rho = R^\alpha_{\rho\nu\mu}B_\alpha,
\end{equation}
which is also equivalent to
\begin{equation}\label{RiemannUpper}
    \nabla_\mu\nabla_\nu B^\rho - \nabla_\nu\nabla_\mu B^\rho = R^\rho_{\alpha\mu\nu}B^\alpha.
\end{equation}
This equivalence can be shown using the following symmetries of the Riemann tensor,
\begin{equation*}
    g_{\beta\alpha}R^\beta_{\mu\nu\rho} = R_{\alpha\mu\nu\rho} = -R_{\mu\alpha\nu\rho} = -R_{\alpha\mu\rho\nu} = R_{\nu\rho\alpha\mu}.
\end{equation*}
The Riemann curvature tensor is the description of the lack of commutativity of covariant derivatives. One may contract the Riemann curvature tensor with itself to form the Ricci curvature tensor,
\begin{equation*}
    R^\alpha_{\rho\alpha\mu} = R_{\rho\mu}.
\end{equation*}By raising an index and contracting again, one forms a scalar quantity, the (Ricci) scalar curvature:
\begin{equation*}
    R_{\rho\mu}g^{\rho\mu} = R^\rho_\rho = R.
\end{equation*}
This process of index raising and contraction is also known as taking the trace of the tensor --- think of taking the trace of some matrix $\mathcal{M}$, the trace is simply $Tr(\mathcal{M}) = \sum^{n}_{i=1}\mathcal{M}_{ii}$.
\begin{definition}A spacetime is flat if and only if the components of the Riemann curvature tensor vanish everywhere.
\end{definition}Since the Riemann tensor is a tensor, if all components vanish everywhere in one coordinate system, they vanish in all coordinate systems.

In order to state the Einstein field equations, we just need one last object, the energy-momentum tensor. This is an object $T^{\mu\nu}$ which we require satisfy:
\begin{align*}
    &\textit{1) Symmetry: } T^{\mu\nu} = T^{\nu\mu}, \\
    &\textit{2) Tensor Transformation: } T'^{\mu\nu}(x') = J^\mu_\rho(x)J^\nu_\sigma(x)T^{\rho\sigma}(x),\\
    &\textit{3) Conservation Law: } \nabla_\mu T^{\mu\nu} = 0.
\end{align*}
Where $J^\mu_\rho (x)$ and $J^\nu_\sigma (x)$ are the Jacobi matrices introduced when transforming a $(2,0)$-tensor \cite{CarrollBook}.
    The energy-momentum tensor describes the matter in spacetime. 
We will work with the electromagnetic energy-momentum tensor defined as \cite{Hawking}
\begin{equation*}
    T^{\mu\nu} = \frac{1}{\mu_0} \left[F^{\mu\sigma}F^\nu_{\,\,\,\,\sigma} - \frac{1}{4}g^{\mu\nu}F_{\sigma\rho}F^{\sigma\rho} \right].
\end{equation*}
This has some interesting properties, and in matrix form for Minkowski space this is
\begin{equation*} T^{\mu\nu} = 
\begin{pmatrix}
\frac{1}{2\mu_0}\left(E^2+B^2\right) & S_1 & S_2 & S_3 \\
S_1 & -\sigma_{11} & -\sigma_{12} & -\sigma_{13} \\
S_2 & -\sigma_{21} & -\sigma_{22} & -\sigma_{32} \\
S_3 & -\sigma_{31} & -\sigma_{32} & -\sigma_{33}
\end{pmatrix}^{\mu\nu}.
\end{equation*}
One may recognise some familiar quantities, $T^{00}$ describes the energy, $T^{0i}$ is the Poynting vector, describing the energy flux, and $\sigma$ is the Maxwell stress tensor. 
Also, $T^{\mu\nu}$ has the property that it is trace-free since $1/(2\mu_0)\left(E^2+B^2\right) -\sum^3_{i=1} \sigma_{ii} = 0$. The Maxwell stress tensor is given explicitly by
\begin{equation*}
    \sigma_{ij} = \frac{1}{\mu_0}\left(E_i E_j - \frac{1}{2} \delta_{ij}E^2 + B_i B_j -\frac{1}{2} \delta_{ij} B^2 \right).
\end{equation*}
We define the Einstein tensor as
\begin{equation*}
    G^{\mu\nu} = R^{\mu\nu} - \frac{1}{2}g^{\mu\nu}R.
\end{equation*}
Finally, the Einstein field equations (without cosmological constant) are
\begin{equation*}
    G^{\mu\nu} = 8\pi G\, T^{\mu\nu}.
\end{equation*}
The left hand side describes the geometry of spacetime and the right hand side describes the matter in the spacetime. 
$G$ is the Newton gravitation constant. This equation can be written in a different way by taking the trace of the equation (contracting with $g_{\mu\nu}$). One then finds $R = -8\pi G T$, thus
\begin{equation*}
    R^{\mu\nu} = 8\pi G\,(T^{\mu\nu}- \frac{1}{2}g^{\mu\nu} T),
\end{equation*}
where $T = T^\alpha_\alpha$ is the trace of the energy-momentum tensor. For our energy-momentum tensor, $T^\alpha_\alpha = 0$ so we arrive at the Maxwell-Einstein equations,
\begin{equation}\label{MaxEin}
    R^{\mu\nu} = 8\pi G\,T^{\mu\nu}.
\end{equation}
\begin{remark}
If the electromagnetic field is so weak that it does affect the curvature of spacetime significantly, it suffices to use Eqs. \eqref{eqn:dFJ} and \eqref{eqn:dF}. However, if the electromagnetic field does meaningfully affect the curvature of spacetime, one must use the full Einstein-Maxwell equations \eqref{MaxEin}.
\end{remark}
\subsection{Electromagnetic Wave Equation in Curved Spacetime}
We can rewrite \eqref{eqn:dFJ} in terms of the four-potential (despite the Faraday tensor being the same in flat and curved spacetimes, we shall now write it explicitly with the covariant derivative for reasons which will become clear later on)
\begin{align*}
    \nabla_\beta F^{\alpha\beta} = \nabla_\beta \nabla^\alpha A^\beta - \nabla_\beta \nabla^\beta A^\alpha = \mu_0 J^\alpha.
\end{align*}
One may again employ the Lorenz gauge in its covariant form,
\begin{equation*}
    \nabla_\mu A^\mu = 0.
\end{equation*}
This, as in the flat case, does not fully specify the gauge condition because one may still transform the vector potential
\begin{equation*}
    \nabla_\mu A'^\mu = \nabla_\mu ( A^\mu + \nabla^\mu \chi) \overset{!} = 0.
\end{equation*}
So, we require the harmonic condition on the scalar function $\chi$ that
\begin{equation*}
    \nabla_\mu \nabla^\mu \chi = 0.
\end{equation*}
Now, in order to use this gauge, the contraction must happen between the vector potential and the first covariant derivative. However, looking at the first term in
\begin{equation} \label{curvedwaveeqn}
    \nabla_\beta \nabla^\alpha A^\beta - \nabla_\beta \nabla^\beta A^\alpha = \mu_0 J^\alpha,
\end{equation}we can commute the two covariant derivatives. Recalling \eqref{RiemannUpper}, this shows how we are able to commute covariant derivatives:
\begin{align*}
    g^{\alpha \sigma}\nabla_\beta \nabla_\sigma A^\beta &= g^{\alpha\sigma} \left(\nabla_\sigma \nabla_\beta A^\beta + R^\beta_{\rho\beta\sigma}A^\rho \right),\\
    &=\nabla^\alpha \nabla_\beta A^\beta + g^{\alpha \sigma} R_{\rho\sigma}A^\rho,\\
    &=\nabla^\alpha \nabla_\beta A^\beta + R_\rho^{\,\,\,\alpha} A^\rho.\\
\end{align*}
Substituting this equation into \eqref{curvedwaveeqn} yields
\begin{equation*}
    \nabla^\alpha \nabla_\beta A^\beta + R_{\rho}^{\alpha} A^\rho - \nabla_\beta \nabla^\beta A^\alpha = \mu_0 J^\alpha,
\end{equation*}which under the Lorenz gauge reduces down to
\begin{equation}\label{WaveEqnCurved}
    - \nabla_\beta \nabla^\beta A^\alpha +  R_{\rho}^{\alpha} A^\rho = \mu_0 J^\alpha.
\end{equation}In flat spacetimes with Minkowski coordinates, the covariant derivatives reduce to partial derivatives with vanishing Ricci tensor so this yields the flat wave equation,
\begin{equation*}
    \partial_\beta \partial^\beta A^\alpha = -\mu_0 J^\alpha.
\end{equation*}
The left hand side of Eq. \eqref{WaveEqnCurved} has a name of its own, the de Rham 'Laplacian' \cite{DeRham} or the de Rham wave operator and one writes this wave equation as
\begin{equation*}
    \left[\Delta_{(dR)}A \right]^\alpha = \mu_0 J^\alpha.
\end{equation*}
A crucial point to notice is that this wave equation has an explicit curvature term, it does not simply follow the minimal coupling method since that would require covariant derivatives to commute -- the Ricci tensor term explicitly comes from this not occurring. The equation we derived requires this term to imply conservation of charge.
We now show a wave solution using the \textit{geometrical optics approximation} in \cite[p.~71]{Wald}.
If one considers a vacuum spacetime, i.e. where $J^\alpha = 0$, one considers a solution to
\begin{equation}\label{eqn:curvewave}
    -\nabla_\beta \nabla^\beta A^\alpha + R^\alpha_\rho A^\rho = 0,
\end{equation}
in situations where the \textit{spacetime scale of variation of the electromagnetic field is much smaller than that of the curvature}, i.e. where spacetime does not curve significantly to affect a wavelike solution. One would then expect to find solutions with nearly constant amplitude,
\begin{equation}\label{eqn: appendixansatzcurved}
    A^\alpha = C^\alpha e^{i k_\mu x^\mu},
\end{equation}
where derivatives of $C^\alpha$ are "small". One then substitutes this Ansatz into \eqref{eqn:curvewave} whilst neglecting the "small" derivatives of $C^\alpha$ and the Ricci curvature term, leaving
\begin{equation} \label{eqn: appendixcurved}
    -\nabla_\beta \nabla^\beta A^\alpha = 0.
\end{equation}
This calculation, reported in Appendix \ref{Appendix: wave curved}, yields the condition that for a non-trivial wave solution, the wave four-vector is lightlike, i.e. $k^\mu k_\mu = 0$.
\subsection{Electrodynamics in Curved Spacetimes}
To achieve the dynamical theory, one must first familiarise oneself with a few concepts. The \textit{worldline} is the trajectory of a particle. In our spacetimes, this is an assignment
\begin{align*}
    &x:\RR \rightarrow \RR^{1,3} \cong \RR^4,\\
    &\lambda \mapsto x \left( \lambda \right) = \begin{pmatrix}
    t \left( \lambda \right) \\
    x^1 \left( \lambda \right) \\
    x^2 \left( \lambda \right) \\
    x^3 \left( \lambda \right)
    \end{pmatrix},
\end{align*}
where $\lambda$ is a parametrisation and is \textit{not} necessarily the (coordinate) time. 
Given a worldline, we can use the metric to classify these in terms of their tangent vectors. We define $t^\mu = \frac{\dd x^\mu (\lambda)}{ \dd\lambda} $ to be the tangent vector to the worldline.
We say that
\begin{align*}
   &t^\mu \text{ is timelike if } g_{\mu\nu} t^\mu t^\nu < 0,\\
    &t^\mu \text{ is lightlike if } g_{\mu\nu} t^\mu t^\nu = 0,\\
    &t^\mu \text{ is spacelike if } g_{\mu\nu} t^\mu t^\nu > 0.
\end{align*}
Particles whose trajectories have timelike tangent vectors travel slower than the speed of light (they have mass); have lightlike tangent vectors travel at the speed of light (they are massless); have spacelike tangent vector travel faster than the speed of light. This can be demonstrated using the \textit{light cone}. The light cone allows the decomposition of spacetime at a point in our spacetime into five sections: those in the future which can be reached by a lightlike worldline; the points in the future which can be reached by a timelike worldline; the points in the past which can be reached by a lightlike worldline; the points in the past which can be reached by a timelike worldline; and the points which can be reached by a spacelike worldline.
One may typically suppress one or two spatial dimensions to ease the drawing of the light cone however the light cone is to represent the four dimensional spacetime.
\begin{figure}[H]
    \begin{center}
\scalebox{0.6}{    
  \begin{tikzpicture}
    \draw [fill=white!60!] (-1.8, -1.8) -- (0, 0) -- (1.8, 1.8);
    \draw [fill=white!60!] (1.8, -1.8) -- (0, 0) -- (-1.8, 1.8);
    \draw (0,1.8) ellipse (1.8cm and 0.1cm);
    \draw (0,-1.8) ellipse (1.8cm and 0.1cm);
    \draw [->] (0,0) -- (1.6,0.7) node [right] {$y$};
    \draw [->] (-2, 0) -- (2, 0) node [right] {$x$};
    \draw [->] (0, -1.8) -- (0, 2.1) node [above] {$t$};
    \node at (0, -1) {};
    \node at (0, -1) [left] {$Past$};
    \node at (2, 2.5){};
    \node at (0, 1.5) [left] {$Future$};
    \node at (3, 2) {};
    \draw[red, thick] (0,0) -- (-1.6,-0.2);
    \draw[blue, thick] (0,0) -- (0.3,1.5);
    \filldraw[black] (1.3,1.3) circle (1pt) node[anchor=west] {B};
    \filldraw[black] (0.3,1.5) circle (1pt) node[anchor=west] {A};
    \filldraw[black] (-1.6,-0.2) circle (1pt) node[anchor=north] {C};
  \end{tikzpicture}
 }
\end{center}
    \caption{Representation of the light cone (suppressing one spatial dimension).}
    \label{fig:lightcone}
\end{figure}
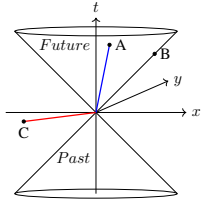
The boundary of the light cone is formed by the lightlike worldlines from the centre point so anything on the light cone's boundary is on a lightlike worldline such as point $B$. Anything inside the light cone (and not on the boundary) is represented by a timelike worldline, point $A$, and anything outside the light cone is on a spacelike worldline, point $C$!
In Newtonian mechanics, the parametrisation is the coordinate time, however in relativistic mechanics because this coordinate time is not invariant under transformations, one must use a so called \textit{proper time}. This is the time experienced by a timelike particle in its frame of reference and this parametrisation is often called $\tau$. Using our convention for metric signature $(-,+,+,+)$, and writing this parametrisation explicitly
\begin{equation*}
    g_{\mu\nu}\,\dd x^\mu \dd x^\nu = \dd s^2 = -c^2 \dd\tau^2,
\end{equation*}
leads to the expression \cite[p. ~9]{CarrollBook}
\begin{equation*}
    \Delta \tau = \int \sqrt{- g_{\mu\nu}\frac{\dd x^\mu}{\dd\lambda}\frac{\dd x^\nu}{\dd\lambda}} \,\dd\lambda.
\end{equation*}
This is the time interval experienced by an observed moving along this timelike worldline. The four-velocity is the normalised tangent vector (living in the future light cone to the worldline of a particle. That is, the particle's four-velocity is
\begin{equation*}
    u^\mu = \frac{d x^\mu}{d \lambda},
\end{equation*}
and if the trajectory is parametrised by its proper time, $\lambda = \tau$, then the four-velocity is normalised, viz., 
\begin{equation*}
    g_{\mu\nu}u^{\mu}u^{\nu} = -1.
\end{equation*}
Therefore, the four-velocity is normalised \textit{if and only if the trajectory is parametrised by its proper time.}
Before writing down the dynamical equation for electromagnetism, we must cast ourselves back to Newtonian mechanics as a motivation. Its dynamics is governed by derivatives with respect to its parameter. In Newtonian mechanics, time is the parameter, however in relativistic mechanics we have a choice of parameter, in particular, we can choose proper time $\tau$. Consider now by the chain rule
\begin{equation*}
    \frac{\dd A^\mu (x(\tau))}{\dd \tau} := \frac{\dd x^\nu (\tau)}{\dd\tau} \partial_\nu A^\mu (x(\tau)).
\end{equation*}
Now, we can apply minimal coupling and define a \textit{covariant parameter derivative},
\begin{align*}
    \frac{\nabla A^\mu (x(\tau))}{\dd\tau} &:= \frac{\dd x^\nu (\tau)}{\dd\tau} \nabla_\nu A^\mu (x(\tau)),\\
    &= \frac{\dd x^\nu (\tau)}{\dd\tau} \left(\partial_\nu A^\mu(x(\tau)) + \Gamma^\mu_{\nu\rho}(x(\tau))A^\rho(x(\tau)) \right),\\
    &= \frac{\dd A^\mu (x(\tau))}{\dd \tau} + \Gamma^\mu_{\nu\rho}(x(\tau))\frac{\dd x^\nu (\tau)}{\dd\tau}A^\rho(x(\tau)).
\end{align*}
This applies to all $(1,0)$-tensors and in particular, to the four-velocity $u^\mu$,
\begin{equation*}
    \frac{\nabla u^\mu (\tau)}{\dd\tau} = \frac{\dd u^\mu (\tau)}{\dd\tau} + \Gamma^\mu_{\nu\rho}(x(\tau))u^\nu(\tau)u^\rho(\tau).
\end{equation*}
In general, the dynamical equation (the generalisation of $F=ma$) is
\begin{equation}\label{f=ma}
     m  \frac{\nabla u^\mu (\tau)}{\dd\tau} = F^{\mu}(x(\tau)),
\end{equation}
where $F^\mu$ is a so-called \textit{four-force}.
Using the concept of a four-velocity parametrised by proper time, one can now write down the dynamical equation for electromagnetism:
\begin{equation*}
    m  \frac{\nabla u^\mu (\tau)}{\dd\tau} = q F^{\mu\nu}u_\nu.
\end{equation*}
The right hand side of the equation describes the Lorentz force, 
$m$ is the mass, and $q$ is the charge of the particle, $F^{\mu\nu}$ is again the Faraday tensor, and $u_\nu = g_{\mu\nu}u^\mu$ is the four-velocity.
\subsection{Geodesics}\label{geodesics}
An important example of \eqref{f=ma} is when $F^\mu = 0$. In Newtonian mechanics, we know that $ma=0$ yields solutions of the form $x = a + bt$, straight lines. These are the shortest distances between two points in Euclidean geometry; the name for the shortest distance is a geodesic. Now, when $F^\mu = 0$, we recover the geodesic equation which is usually written in terms of the worldlines
\begin{equation}\label{geodesic equation}
    \frac{\dd^2 x^\mu(\tau)}{\dd \tau^2} + \Gamma^\mu_{\nu\rho} \frac{\dd x^\nu(\tau)}{\dd\tau}\frac{\dd x^\rho(\tau)}{\dd\tau} = 0.
\end{equation}
This plays an important role in differential geometry as it tells you the shortest path between two points. An example would be if one used the metric for a sphere of radius 1,
\begin{equation*}
    \dd s^2 = \dd\theta^2 + \sin^2{\theta}\dd\varphi^2,
\end{equation*}
solving this equation would yield solutions of great circles --- circles that pass through the centre of the sphere, diving it into two equal parts. Solving this equation is in general not easy.
The geodesic equation is more familiar to the reader than at first glance, it is actually the Euler-Lagrange equation for two different action integrals. Classically, the dynamics of a system described by a Lagrangian $\mathcal{L}[q(t), \dot{q}(t),t]$ can be found by looking at the minimum of its action. The action $S[q(t)]$ is defined as an integral
\begin{equation*}
    S[q(t)] = \int \mathcal{L}[q(t), \dot{q}(t),t] \dd t.
\end{equation*}
In Newtonian mechanics, the parameter used is time. In relativity, time is on an equal footing to space so we cannot single it out and integrate over it, therefore we integrate over our parameter $\lambda$. The two actions relevant to the geodesic equation are
\begin{align*}
    S_1 = \int \sqrt{-g_{\mu\nu}\frac{\dd x^\mu}{\dd\lambda}\frac{\dd x^\nu}{\dd\lambda}}\dd\lambda, \text{ and }
    S_2 = \frac{1}{2}\int g_{\mu\nu} \frac{\dd x^\mu}{\dd\lambda}\frac{\dd x^\nu}{\dd\lambda} \dd\lambda.
\end{align*}
The first has a geometric meaning since the integrand, $\sqrt{\dd s^2}=\dd s$, is the line element. 
Therefore, by computing the Euler-Lagrange equations minimizes the path. The second is however computationally easier to work with.
\subsection{Flat Dipole Radiation}
For fullness of this review, we present the flat spacetime tensorial form of dipole radiation. Dipole approximations are known in curved spacetimes, however these are more involved and for compactness, we do not cover this. For the case of Schwarzschild, see \cite{CurvedDipole} which utilises the Newman-Penrose formalism.

Consider the case of electric dipole radiation by imagining two small metal spheres, connected by a thin wire, separated by a distance $d$, with a charge oscillating between the two balls, $q(t) = q_0 \cos(\omega t)$,
where $\omega$ is the angular frequency and the maximum value of the dipole moment is $p_0 = q_0 d$.
Following \cite[p.~469-471]{Griffiths}, the scalar and vector potentials can be approximated by the \textit{electric dipole approximation} in cylindrical coordinates, which reads
\begin{align*}
    \phi &= -\frac{\mu_0 p_0 \omega}{4\pi r} \cos{\theta}\sin{\left(\omega\left(t-r\right)\right)}, \\
    \bm{A} &= -\frac{\mu_0 p_0 \omega}{4\pi r}\sin{\left(\omega\left(t-r\right)\right)}\hat{\bm{z}},
\end{align*} 
or, in compact tensorial notation
\begin{equation*}
    A^\mu = -\frac{\mu_0 p_0 \omega}{4\pi r}\sin{\left(\omega\left(t-r\right)\right)}
    \begin{pmatrix}
    \cos{\theta} \\
    0\\
    0\\
    1
    \end{pmatrix}.
\end{equation*}
One should note that in the argument of the trigonometric functions, $t - r = t - |x-x'| = t_r$, the retarded time.
Then using $\hat{\bm{z}} = \cos{\theta}\,\hat{\bm{r}} - \sin{\theta}\,\hat{\bm{\theta}}$, one can calculate the fields that these potentials create, which reads
\begin{align*}
    \bm{E} &= -\frac{\mu_0 p_0 \omega^2}{4\pi r} \sin{\theta}\cos{\left(\omega\left(t-r\right)\right)} \hat{\bm{\theta}}, \\
    \bm{B} &= -\frac{\mu_0 p_0 \omega^2}{4\pi r} \sin{\theta}\cos{\left(\omega\left(t-r\right)\right)} \hat{\bm{\varphi}}.
\end{align*}The Faraday tensor here is
\begin{equation*}
    F^{\mu\nu} = -\frac{\mu_0 p_0 \omega^2}{4\pi r} \sin{\theta}\cos{\left(\omega\left(t-r\right)\right)}
    \begin{pmatrix}
    0 & 0 & 1 & 0 \\
    0 & 0 & 1 & 0 \\
    -1 & -1 & 0 & 0\\
    0 & 0 & 0 & 0
    \end{pmatrix}.
\end{equation*}
These fields correspond to waves of frequency $\omega$, travelling radially at the speed of light. 
The case of dipole radiation is considered in \cite[p. ~271]{Jackson} with $q(t) = q_0 \ee^{-\ii\omega t}$,
leading to the complex fields
\begin{align*}
    \bm{E} = -\frac{\mu_0 p_0 \omega^2}{4\pi r} \sin{\theta}\,e^{i\omega\left(r-t \right)} \hat{\bm{\theta}},\,
    \bm{B} = -\frac{\mu_0 p_0 \omega^2}{4\pi r} \sin{\theta}\,e^{i\omega\left(r-t \right)} \hat{\bm{\varphi}}.
\end{align*}
The strength of this notation is that it allows one to recognise that both $\bm{E}$ and $\bm{B}$ are spherical waves of the form $u(r,t)=  1/r\,F(r-t) + 1/r\,G(r+t)$.
\begin{figure}[H]
    \centering
    \includegraphics[width=9 cm]{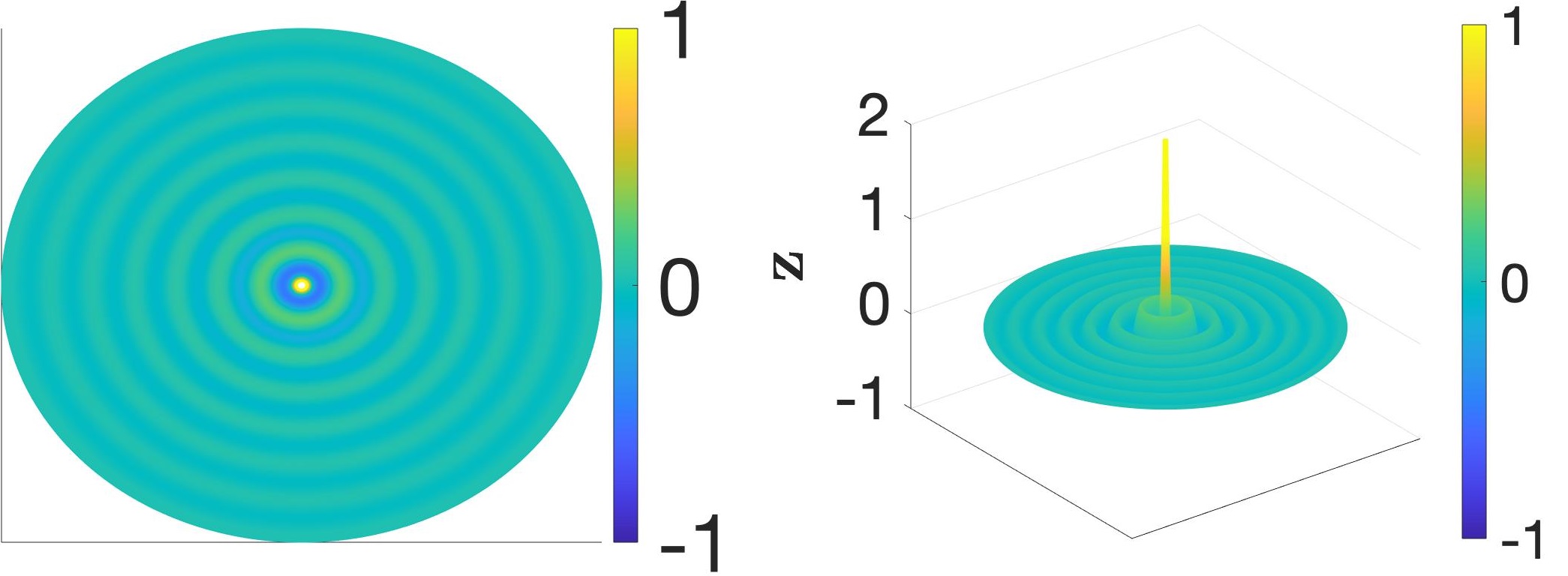}
    \caption{Spherical Wave with $\theta = \pi/2$, from above and from the side.}
    \label{fig:spherical1}
\end{figure}
As shown in Fig. \ref{fig:spherical1}, as $r \rightarrow \infty$ increases, the waves become weaker in amplitude and 
tend to plane waves as $r \rightarrow \infty$ 
\subsection{Propagation of Radiation in Curved Spacetimes}
We have seen that an oscillating dipole emanates spherical electromagnetic waves and it is known that electromagnetic waves are propagated by the photon which travels on lightlike worldlines. Another term for lightlike is \textit{null}. Light and, more generally, photons travel on \textit{null geodesics}. Geodesics were described in section \ref{geodesics}; the \textit{null} part refers to the lightlike condition. Consider the general metric $\dd s^2 = g_{\mu\nu}\dd x^\mu \dd x^\nu$. This would be null when $\dd s^2 = 0$. Therefore, one considers the ray paths under this condition.
A useful geometry is the Schwarzschild geometry, described by the metric
\begin{equation*}
    \dd s^2 = -\left( 1- \frac{2GM}{r}\right) \dd t^2 + \frac{\dd r^2}{1-\frac{2GM}{r}} + r^2 \dd\Omega^2,
\end{equation*}
which describes how a spherically symmetric object (such as a perfectly spherically symmetric star) curves the spacetime about it. $M$ is a mass parameter, related to the Newtonian mass by $M = GM_{\text{Newt}}$, where $G$ is the gravitational constant and $M_{\text{Newt}}$ is the Newtonian measured mass. One notes that the case where $M = 0$ gives the Minkowski metric of special relativity so this encompasses the case of no mass. The spherical symmetry is described by $\dd\Omega^2 = \dd\theta^2 + \sin^2 \theta \dd\varphi^2$, the metric for the sphere\footnote{This is the definition of $\dd\Omega^2$ in this work, however it is important to know that this is also used for spherical symmetry in higher dimensions. More explicitly, the dimension of our spacetime is $4$: we use $1$ for time, $1$ for radial distance and $2$ for angles. One also considers a spacetime of dimension $n$, then $\dd\Omega^2$ would then be the metric of $S^{n-2}$, the sphere of dimension $(n-2)$.}. Though, this is not the most general form of this metric
\begin{equation}\label{general spherical metric}
    \dd s^2 = -F(r) \dd t^2 + \frac{\dd r^2}{F(r)} + r^2 \dd\Omega^2,
\end{equation}
which encompasses metrics such as the Reissner-Nordstr\"om metric describing a non-rotating charged mass and also the anti-de Sitter spacetime metric, a spacetime with constant negative curvature - its Ricci scalar is constant and negative. Such a spacetime is of interest for string theorists thanks to its role in the AdS/CFT correspondence \cite{Maldacena_1999}.

When studying the geodesics of \textit{null rays}, one considers both \textit{radial}, and \textit{non-radial} null rays. The radial case is when there is no motion within the angles, and the non-radial case is when one does allow movement within the angles.
\subsubsection{Radial Null Rays}
One has $\dd s^2 = 0$ for the null condition and the radial condition comes into play in the Euler-Lagrange equations for the \textit{geodesic Lagrangian}. From section \ref{geodesics}, one has
\begin{align*}
    S &= \int \frac{1}{2}\left( -F(r) \dot{t}^2 + \frac{1}{F(r)}\dot{r}^2 + r^2 \dot{\theta}^2 + r^2 \sin^2{\theta} \dot{\varphi}^2 \right)\dd\lambda .
\end{align*}
Since these rays are purely radial one has $\dot{\theta} = \dot{\varphi} = 0$ where a dot refers to differentiation with respect to $\lambda$. The \textit{geodesic Lagrangian} is then
\begin{equation*}
    L = \frac{1}{2}\left( -F(r) \dot{t}^2 + \frac{1}{F(r)}\dot{r}^2 \right).
\end{equation*}
To refresh the reader, the Euler-Lagrange equations are
\begin{equation*}
    \frac{\dd}{\dd\lambda}\frac{\partial L}{\partial \dot{x}^\mu} = \frac{\partial L}{\partial x^\mu}.
\end{equation*}
One can see that the geodesic Lagrangian is independent of $t$, therefore $\partial L/\partial t = \textit{const}$, and we call this constant $-E$. More explicitly, by setting $\mu = 0$, we have,
\begin{align*}
    \underbrace{\frac{\dd}{\dd\lambda}\frac{\partial L}{\partial \dot{x}^0} = \frac{\dd}{\dd\lambda}p_t}_{\text{LHS of Euler-Lagrange Equations}} = \underbrace{\frac{\partial L}{\partial x^0} = \frac{\partial L}{\partial t} =0.}_{\text{RHS of Euler-Lagrange Equations}}
\end{align*}$p_t$ is the momentum conjugate to the time $t$ in the context of Lagrangian mechanics. Taking the derivative of $L$ with respect to $\dot{t}$, we have
\begin{equation*}
    p_t = -F(r) \dot{t} = -E.
\end{equation*}
Using the notation, $\dd t = E/F(r) \,\dd\lambda$, then substituting into our metric, we have
\begin{align}
    0 &=-F(r) \frac{E^2}{F^2(r)}\dd\lambda^2 + \frac{\dd r^2}{F(r)} + r^2 \dd\Omega^2,\\
    \implies 0 &=-F(r) \frac{E^2}{F^2(r)} + \frac{1}{F(r)} \dot{r}^2 + r^2\left(\dot{\theta} + \sin^2\theta\,\dot{\varphi} \right), \\
    &= \frac{1}{F(r)}\left( -E^2 + \dot{r}^2\right).
\end{align}This yields our second condition on our radial null rays,$\dot{r} = \pm E$. This clearly has the solution $r = \pm E \lambda + const$.
One can solve for $t$ in terms of $r$ by realising that 
\begin{equation*}
    \frac{\dot{t}}{\dot{r}} =\frac{\dd t}{\dd\lambda} \frac{\dd\lambda}{\dd r} = \frac{\dd t}{\dd r} = \mp \frac{1}{F(r)}.
\end{equation*}
Then one solves for this ODE to see
\begin{equation*}
    \pm t = r + 2GM \, \ln{\left(\frac{r}{2GM}-1 \right)}+const.
\end{equation*}
These were derived from the geodesic Lagrangian using the Euler-Lagrange equations so we can clearly see that these are geodesics but it is not always so easy to see that these are still null (lightlike).
\subsubsection{Non-Radial Null Rays}
By the spherical symmetry of the geometry, one choose coordinates such that, for example, $\theta = \pi/2$ since a geodesic is described in an invariant plane \cite[p. ~11]{Thesis}. Thus $\dd\theta = 0$, $\dot{\theta} = 0$, and $\sin{\theta} = 1$. This simplifies the metric greatly,
\begin{equation*}
    \dd s^2 = -F(r) \dd t^2 + \frac{\dd r^2}{F(r)} + r^2 \dd\varphi^2.
\end{equation*}In similar fashion to before, the geodesic Lagrangian in this case is
\begin{equation*}
    L = \frac{1}{2} \left(-F(r) \dot{t}^2 + \frac{1}{F(r)} \dot{r}^2 + r^2 \dot{\varphi}^2 \right).
\end{equation*}Upon inspection, one sees that the Lagrangian is independent of both $t$ and $\varphi$, leading to one of the same conservation laws as before $F(r)\,\dot{t} = E$, as well as
\begin{align} \label{ell}
    p_\varphi = \frac{\partial L}{\partial \dot{\varphi}} = r^2 \dot{\varphi} = \ell,
\end{align}
which describes conservation of angular momentum.
Returning to the metric,
\begin{align*}
    0 = ds^2 &= -F(r)dt^2 + \frac{1}{F(r)}dr^2 + r^2 d\varphi^2, \\[-4pt]
    \implies 0 &=-F(r) \dot{t}^2 + \frac{1}{F(r)}\dot{r}^2 + r^2 \dot{\varphi}^2, \\[-4pt]
    &= -\frac{E^2}{F(r)} + \frac{\dot{r}^2}{F(r)} + \frac{\ell^2}{r^2}, \\[-4pt]
   \implies \dot{r}^2 & = E^2 - \frac{\ell^2}{r^2}F(r),\\[-4pt]
\end{align*}
leaving the radial equation ODE,
\begin{equation*}
   {\dot{r}= \pm \sqrt{ E^2 - \frac{\ell^2}{r^2}\left(1- \frac{2GM}{r} \right)}}.
\end{equation*}
This same equation can be found \cite[p. ~1]{EscapeofPhotons} with differing constants, we have $\alpha = \ell$, and $\beta =E/\ell$. However, if one is simply interested in the orbit that the ray takes, one looks at $\dd\varphi/\dd r = \dot{\varphi}/\dot{r}$.

\section{Gravitational Redshift}\label{sec:: grav redshift}
The effect of a mass curving spacetime causes a shift in the frequency of light/electromagnetic radiation emitted by one particle observed by another and this has applications within communication, such as two satellites communicating. We present here a geometric derivation based on Schr\"odinger's derivation \cite[p.~47-53]{Redshift}.
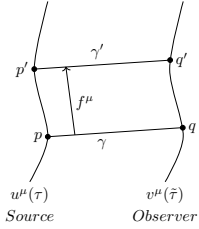
\begin{figure}[H]
    \centering
\scalebox{0.6}{
\begin{tikzpicture}
    \draw plot [smooth] coordinates {(0,0) (0.4,1) (0.1,2.5) (0.4,4)};
    \draw plot [smooth] coordinates {(3,0) (3.4,1) (3.1,2.5) (3.4,4)};
    \draw (0.4,1)--(3.36,1.2);
    \filldraw[black] (0.4,1) circle (1.5pt) node[anchor=east] {$p$};
    \filldraw[black] (3.39,1.2) circle (1.5pt) node[anchor=west] {$q$};
    \draw (0.1,2.5) -- (3.14,2.7);
    \filldraw[black] (0.1,2.5) circle (1.5pt) node[anchor=east] {$p'$};
    \filldraw[black] (3.11,2.7) circle (1.5pt) node[anchor=west] {$q'$};
    \node at (0, 0) [below] {$u^\mu (\tau)$};
    \node at (3, 0) [below] {$v^\mu (\tilde{\tau})$};
    \node at (0, -1) [above] {$Source$};
    \node at (3, -1) [above] {$Observer$};
    \draw [->](1,1.04) -- (0.8,2.55);
    \node at (0.9,1.755) [right] {$f^\mu$};
    \node at (1.5,2.6) [above] {$\gamma'$};
    \node at (1.6,0.6)[above] {$\gamma$};
\end{tikzpicture}
}
    \caption{Two observers and their respective worldlines, interacting via emission of photons.}
    \label{fig:redshift}
\end{figure}
Consider Fig. \ref{fig:redshift} above. This shows two particles travelling along timelike worldlines, parametrised by their own proper times, $\tau$ and $\tilde{\tau}$, each having their own four-velocities, $u^\mu (\tau)$ and $v^\mu(\tilde{\tau})$. Naming the left hand side particle the source and the right hand side the observer, consider when the source is at $p$ along its worldline and it emits a photon that the observer observes at point $q$. Then, (very) shortly after, the source emits a second photon at point $p'$, being observed at point $q'$. The world lines for each photon are
\begin{align*}
    \gamma &= x^\mu (\lambda),\\
    \gamma' &= y^\mu (\lambda) = x^\mu (\lambda) + f^\mu (\lambda),
\end{align*} where both $\gamma$ and $\gamma'$ are null geodesics and $f^\mu$ is "small".
From $p$ to $p'$, the source moves a distance of $\dd\tau (u^\mu)_p = (f^\mu)_p$ and the observer moves a distance of $ \dd\tilde{\tau} (v^\mu)_q = (f^\mu)_q$ where $\dd\tau$ is the proper time elapsed for the source between $\gamma$ and $\gamma'$ and $\dd\tilde{\tau}$ is similarly defined.
It is shown in "Proof That $f_\mu\dot{x}^\mu$ Is Constant Across $\gamma$" that the quantity $f_\mu \dot{x}^\mu$ is constant over $\gamma$. As a corollary of this, and writing $\dot{x} = \dd x/\dd \tau$, we have
\begin{align*}
    (\dot{x}^\mu f_\mu)_p &= (\dot{x}^\mu f_\mu)_q,\\
    (\dot{x}^\mu u_\mu)_p \dd\tau & = (\dot{x}^\mu v_\mu)_q \dd\tilde{\tau},\\
    \implies \frac{\dd\tau}{\dd\tilde{\tau}} &= \frac{(\dot{x}^\mu v_\mu)_q}{(\dot{x}^\nu u_\nu)_p},
\end{align*} where $\dot{x}^\mu$ is the tangent of the null geodesic that the photon takes and one might call this instead $w^\mu$. Calling $\nu$ the frequency of the observed light, we now have
\begin{equation}\label{eqn: Redshift}
    \frac{\nu_\text{observed}}{\nu_\text{emitted}} = \frac{\left(w^\mu v_\mu \right)_q}{\left(w^\sigma u_\sigma \right)_p},
\end{equation} our gravitational redshift formula.

In our calculations, we use only spacetimes whose metrics are Schwarzschild-like, i.e. of the form ($16$), describing well the exterior regions of spherical objects such as planets, stars, or non-rotating black holes. However, not all spacetime metrics are of this form. A ubiquitous cosmological model of the universe which encompasses its expansion is the Friedmann-Lemaître-Robertson-Walker (FLRW) model, with metric
\begin{equation*}
\dd s^2 = -\dd t^2 + a^2(t)\left(\dd x^2+\dd y^2+\dd z^2 \right),
\end{equation*}
where $a(t)$ is referred to as the scale factor of the universe. The specific details of the universe it describes are of a homogeneous, isotropic, expanding universe. This model is explored in \cite{Budko}, where the resulting electric field to the sourced Maxwell equations is found. By introducing the variable $\tilde{t} = \int_{t_0}^t 1/a(t') \dd t'$ and noting $\dd\tilde{t} = 1/a(t) \dd t$ via the fundamental theorem of calculus, the metric is transformed into
\begin{equation*}
\dd s^2 = a^2(t)(-\dd\tilde{t}^2 + \dd x^2 + \dd y^2 + \dd z^2) = a^2(t) \dd s_{\text{Minkowski}}^2.
\end{equation*}
Since the Minkowski metric and FLRW metric differ by only an overall positive scale factor, they are said to be conformally equivalent. The redshift of monochromatic light in FLRW spacetime is calculated in \cite{Budko} for the case $a(t) = a_0 + bt$ as
\begin{equation*}
\frac{\nu_{\text{observed}}}{\nu_{\text{emitted}}} = \ee^{-b|\bm{x}-\bm{x}'|},
\end{equation*} where $\bm{x}$ and $\bm{x}'$ are the positions of the observer and source respectively.
\subsection{Stationary Observers}
Consider the case where both the source and observer are separated only radially and neither are moving spatially. The light being emitted from the source arrives at the observer and travels along a radial, null geodesic so that we have explicit formulae for the elements of the tangent vector of this geodesic. Choosing an energy scale such that $E = 1$, one has
\begin{equation*}
    w^\mu = (\dot{t}, \dot{r}, 0 ,0) = \left(\frac{1}{F}, \pm1, 0, 0\right). \footnote{In $\dot{r}$, we have two possibilities for the cases whether the light's $r$-coordinate is increasing or decreasing.}\footnote{Note also since $\dot{\varphi}=0$, that $\ell = 0$, rendering $\dot{r}=\pm1$.}
\end{equation*} (In $\dot{r}$, we have two possibilities for the cases whether the light's $r$-coordinate is increasing or decreasing. Also note that since $\dot{\varphi}=0$, that $\ell = 0$, rendering $\dot{r}=\pm1$.) The four-velocities of the source and observer also take simple forms, simply being $u^\mu = (u^0, 0,0,0)$ and $v^\mu = (v^0,0,0,0)$. We have one final piece of information to use --- since the four-velocities are parametrised by proper time, each is normalised, thus one has that
\begin{align*}
    -1 = u^\mu u_\mu = -F (u^0)^2
    \implies u^0 = \frac{1}{\sqrt{F}}.
\end{align*}The positive sign is chosen since the four-velocity points towards future regarding time; in terms of the lightcone, the four-velocity lives in the future null cone rather than the past. We can now use the redshift formula \eqref{eqn: Redshift} to calculate what the redshift due to gravity is.
Using the notation $F_i = 1- 2GM/r_i$ where the source is located at $r_1$ and the observer at $r_2$,
\begin{align*}
    \frac{\nu_\text{observed}}{\nu_\text{emitted}} = \frac{\left(w^\mu v_\mu \right)_q}{\left(w^\sigma u_\sigma \right)_p}
    = \frac{w^0v_0}{w^0u_0}
    = \frac{\frac{1}{F_2}(-F_2\frac{1}{\sqrt{F_2}})}{\frac{1}{F_1}(-F_1\frac{1}{\sqrt{F_1}})}
    =\sqrt{\frac{F_1}{F_2}}.
\end{align*}
Thus the final formula is given by
\begin{equation}\label{eqn:shift}
   \frac{\nu_\text{observed}}{\nu_\text{emitted}} = \sqrt{\frac{1-\frac{2GM}{r_1}}{1-\frac{2GM}{r_2}}},
\end{equation}
showing the gravitational redshift. 
Then by rearranging and observing at infinity, one has
\begin{align}
    \nu_\text{observed} &= \sqrt{1-\frac{2GM}{r_1}} \,\nu_\text{emitted}, \label{eqn standard redshift}\\
    &\approx \left(1 - \frac{GM}{r_1}\right) \nu_\text{emitted}.\label{eqn:redshiftbook}
\end{align}This approximation is often found in textbooks to describe gravitational redshift. This formula can be found in, for example \cite[p. ~620]{CarrollAstronomy} by noting $\nu_\text{observed} = v_\infty$ and $\nu_\text{emitted} = v_0$.
\subsubsection{Radial Motion}
Now we have the groundwork set, one can introduce the case where both the source and observer are moving radially. The change to the setup is in the four-velocity of the source and observer. We now have $u^\mu = (u^0, u^r, 0, 0)$ and $v^\mu = (v^0, v^r, 0, 0)$ such that $u^\mu u_\mu = v^\mu v_\mu = -1$, i.e. that $u$ and $v$ are timelike. This translates to 
\begin{equation*}
    -F_1 (u^0)^2 + \frac{1}{F_1} (u^r)^2 = -1.
\end{equation*}where we want to solve for the zeroth component. Solving and simplifying, we therefore have
\begin{equation*}
    u^0 = \frac{1}{F_1} \sqrt{F_1 + (u^r)^2},
\end{equation*}
with the same calculation being made for $v^0$.
The redshift formula now tells us that
\begin{align*}
    \frac{\nu_\text{emitted}}{\nu_\text{observed}} &= \frac{w^0u_0 + w^1 u_1}{w^0 v_0 + w^1 v_1}
    =\frac{\frac{1}{F_1}(-F_1 u^0) \pm \frac{1}{F_1}u^r}{\frac{1}{F_2}(-F_2 v^0) \pm \frac{1}{F_2}v^r},\\
    &=\frac{F_2}{F_1}\left(\frac{F_1u^0 \mp u^r}{F_2 v^0 \mp v^2}\right)
    =\frac{F_2}{F_1} \left(\frac{\sqrt{F_1 + (u^r)^2}\mp u^r}{\sqrt{F_2 + (v^r)^2}\mp v^r} \right).
\end{align*}Filling out this equation, one retrieves
\begin{equation}\label{eqn:shiftr}
    \frac{\nu_\text{emitted}}{\nu_\text{observed}} = \frac{1-\frac{2GM}{r_2}}{1-\frac{2GM}{r_1}}\left( \frac{\sqrt{1-\frac{2GM}{r_1} + (u^r)^2}\mp u^r}{\sqrt{1-\frac{2GM}{r_2} + (v^r)^2}\mp v^r} \right),
\end{equation}for the redshift due to both radial motion and a gravitational field described by the Schwarzschild metric. 
Again, under the stationary assumption with $r_2$ at infinity, we recover \eqref{eqn standard redshift}. Taking the Newtonian limit of this formula, i.e. such that gravitational effects are negligible $F_i \approx 1$, we use $r_2$, $r_1\gg2GM$, $u^r$, $v^r\ll1$, and keep to leading order in $u^r$ and $v^r$, we find $\nu_{\text{emitted}}/\nu_{\text{observed}} = (1\mp u^r)/(1\mp v^r)$. This is exactly the classical Doppler effect. This calculation tells us that the $\mp$-sign terms in \eqref{eqn:shiftr} give the leading Doppler contributions whereas the $F_i$ terms tell us about the gravitational contributions.

\subsubsection{Radial and Azimuthal Motion}
We can also go one step further with our calculations and allow azimuthal motion of our source and observer. We will use units such that $c=1$, $E=1$, and $\ell = 1$.\footnote{We can do this as long as $\dot{\varphi}>0$, since we can form dimensions of time, mass, and length using $E$, $\ell$, and $c$.} To do this, we require that our light travels on a null, non-radial geodesic given by
\begin{equation*}
    w^\mu = (\dot{t},\dot{r},0,\dot{\varphi}) = \left(\frac{1}{F}, \pm\sqrt{1-\frac{1}{r^2}F}, 0, \frac{1}{r^2}\right), 
\end{equation*} and that our four-velocities are of the forms \newline $u^\mu = (u^0, u^r, 0, u^\varphi)$ and $v^\mu = (v^0, v^r, 0, v^\varphi)$. We write out explicitly that $u_\mu = (-F_1 u^0, \frac{1}{F_1}u^r, 0, r_1^2 u^\varphi)$ to be clear on the lowered index form of the four-velocity. Again, the four-velocity must be normalised such that
\begin{align*}
    u^\mu u_\mu = -F_1 (u^0)^2 + \frac{1}{F_1} (u^r)^2 + r_1^2 (u^\varphi)^2 = -1,
\end{align*}giving
\begin{equation*}
    u^0 = \frac{1}{F_1} \sqrt{F_1 + (u^r)^2 + r_1^2 F_1(u^\varphi)^2}.
\end{equation*}
The redshift formula gives us
\begin{equation*}
    w^\mu u_\mu = -u^0 \pm \frac{1}{F_1}u^r \sqrt{1-\frac{1}{r_1^2}F_1 } + u^\varphi,
\end{equation*}and
\begin{equation*}
    w^\mu v_\mu = -v^0 \pm \frac{1}{F_2}v^r \sqrt{1-\frac{1}{r_2^2}F_2 } + v^\varphi.
\end{equation*}Taking the quotient and substituting in for $u^0$, $v^0$, one sees
\begin{align}\label{eqn:shiftphi}
   &\frac{\nu_\text{emitted}}{\nu_\text{observed}} = 
   \frac{F_2}{F_1}\times \nonumber \\ 
   &\left(\frac{\sqrt{F_1+(u^r)^2 + r_1^2F_1(u^\varphi)^2}\mp u^r\sqrt{1-\frac{1}{r_1^2}F_1 } - F_1 u^\varphi} {\sqrt{F_2+(v^r)^2 + r_2^2F_2(v^\varphi)^2}\mp v^r\sqrt{1-\frac{1}{r_2^2}F_2 } -F_2 v^\varphi} \right),
\end{align}
giving the equation for the redshift due to gravity and both radial and azimuthal motion. 

Taking the non-azimuthal limit requires care as the units with $\ell = 1$ can no longer hold. Therefore, to do this, one must first place $\ell$ back into the equation and then substitute $u^\varphi = 0$, $v^\varphi = 0$, $\ell = 0$, which recovers Eqn. \eqref{eqn:shiftr}.

Together, Eqs. \eqref{eqn:shift}, \eqref{eqn:shiftr}, and \eqref{eqn:shiftphi} give the observed redshifts for different set ups between observer and source, and are relevant wherever the Schwarzschild geometry, or a Schwarzschild-like geometry, can describe a spacetime such as outside the Schwarzschild radius of a Schwarzschild black hole, or outside the surface of a spherically symmetric star. The equations show the combined effects of both gravitational redshift and the Doppler shift due to the motion of both source and observer.

We can see the well-known gravitational redshift by plotting \eqref{eqn:shift}. We assume that $r_2>r_1>2GM$ and that the observer at $r_2$ is not moving. Gravitational effects will be strongest close to the mass. With appropriate units, we can normalise $2GM = 1$. We then look at the region $1<r_1<3/2$, that is $r_2 = 3/2$. Since in most physical situations, both source and observer will be far away from the Schwarzschild radius, the effect of the gravitational redshift is negligible, however it is still of interest for purposes of reviewing phenomena arising from the general relativistic formalism.

\begin{figure}[H]
    \centering
    \scalebox{1}{\includegraphics[width=8cm]{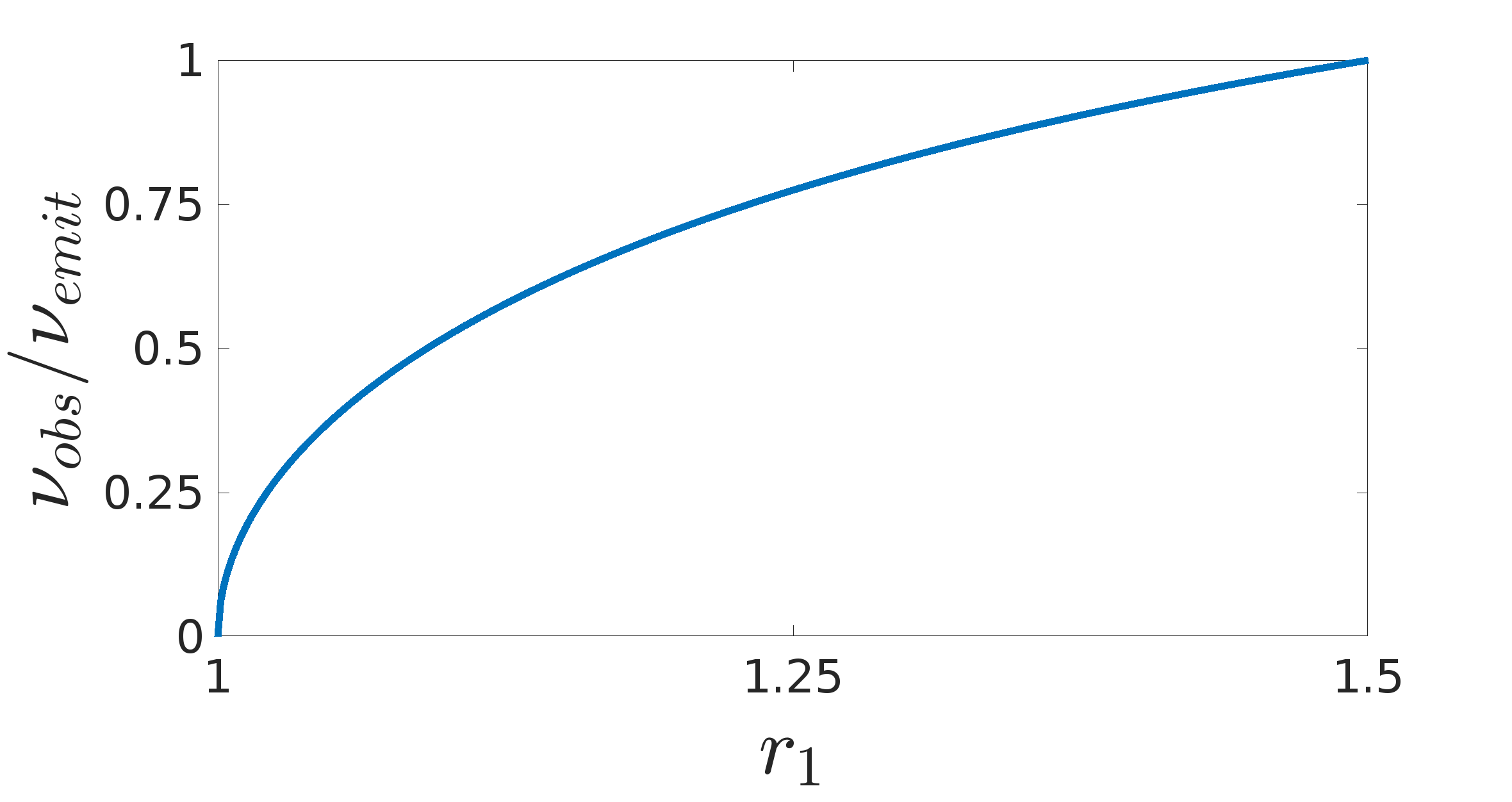}}
    \caption{Gravitational Redshift.}
    \label{fig:dopplerplots}
\end{figure}
Figure \eqref{fig:dopplerplots} is zero at $r_1=1$, which physically represents the Schwarzschild radius. This radius is the point at which all future-directed non-spacelike curves must go inwards to the mass at the centre. At this point, the frequency is infinitely redshifted, hence no frequency can be observed. At $r_1 = r_2 = 3/2$, we find $\nu_{obs}/\nu_{emit} = 1$ as expected.

\section{Conclusion}
We reviewed the main relevant concepts of flat spacetime electromagnetism in order to introduce the general relativistic electromagnetism. We apply this formalism to derive two equations determining the redshift of light detected by a moving observer, as emitted by a moving observer. These equations couple together the effects of the Doppler shift and gravitational redshift to give a single equation, depending on the geometry of the two moving observers. Further research in this area would be in the direction of how interference in curved spacetime might affect communication between two antennas.

\appendices
\section{Geometric Optic Wave Equation Solution}
\label{Appendix: wave curved}
Substituting Eq. \eqref{eqn: appendixansatzcurved} into \eqref{eqn: appendixcurved}, we find
\begin{align*}
    0 &= -\nabla_\beta \nabla^\beta A^\alpha,\\ 
    &= -\nabla_\beta \nabla^\beta \left( C^\alpha \ee^{\ii k_\mu x^\mu} \right).
\end{align*}
The covariant derivative on a scalar is just a partial derivative and note that $\partial_\mu x^\nu = \partial x^\nu / \partial x^\nu = \delta^\mu_\nu$. Applying this twice yields
\begin{align*}
    &= -C^\alpha \nabla_\beta \left(\ii k_\mu \ee^{\ii k_\sigma x^\sigma} \partial^\beta x^\mu\right), \\
    &= - \ii C^\alpha \, k_\mu \delta^{\beta \mu} \, \nabla_\beta \ee^{\ii k_\sigma x^\sigma},\\
    &= -\ii C^\alpha \, k_\mu \delta^{\beta \mu} \, \left(\ii k_\sigma \ee^{\ii k_\rho x^\rho} \partial_\beta x^\sigma \right).
\end{align*}
Simplifying by using the Kronecker deltas to change indices and the metrics to raise or lower indices, we find
\begin{align*}
   -\ii C^\alpha \, k_\mu \delta^{\beta \mu} \, \left(\ii k_\sigma \ee^{\ii k_\rho x^\rho} \partial_\beta x^\sigma \right) &=C^\alpha \,\delta^{\beta \mu} \, \delta^\sigma_\beta \, k_\mu k_\sigma \, \ee^{\ii k_\rho x^\rho},\\
    &=C^\alpha e^{\ii k_\rho x^\rho} \, g^{\beta \nu}\delta^\mu_\nu \,  \delta^\sigma_\beta \, k_\mu k_\sigma,\\
    &= A^\alpha g^{\beta \nu}\delta^\sigma_\nu k_\sigma \, \delta^\mu_\nu k_\mu,\\
    &=A^\alpha g^{\beta \nu}k_\nu k_\beta,\\
    &= A^\alpha k^\nu k_\nu.
\end{align*}
\section{Proof that $f_\mu \dot{x}^\mu$ is constant over $\gamma$}\label{ProofTheorem}
Given that $\gamma'$ is null, we have $0 = g_{\mu\nu}(y) \dot{y}^\mu \dot{y}^\nu$. By Taylor expansion to first order in $f$,
\begin{align*}
    0&= \left(g_{\mu\nu}(x) + f^\sigma\partial_\sigma g_{\mu\nu}(x) \right)\left(\dot{x}^\mu + \dot{f}^\mu \right)\left( \dot{x}^\nu + \dot{f}^\nu \right),\\
    &=g_{\mu\nu}\dot{x}^\mu \dot{x}^\nu + \dot{x}^\mu \dot{x}^\nu \left( \partial_\sigma g_{\mu\nu}\right) f^\sigma + 2 g_{\mu\nu}\dot{x}^\mu\dot{f}^\nu.
\end{align*}
We also have $g_{\mu\nu}\dot{x}^\mu\dot{x}^\nu = 0$ because $\gamma$ is null, and also that
\begin{equation*}
    2g_{\mu\nu} \dot{x}^\mu \dot{f}^\nu = 2 \frac{\dd}{\dd\lambda} \left( g_{\mu\nu} \dot{x}^\mu f^\nu \right) - 2 f^\sigma \frac{\dd}{\dd\lambda}\left( g_{\mu\sigma}\dot{x}^\mu \right).
\end{equation*}Thus returning to the above equality and continuing, we have
\begin{align*}
    0 & = 2\frac{\dd}{\dd\lambda} \left(f_\mu \dot{x}^\mu \right) - 2 f^\sigma \left(\frac{\dd}{\dd\lambda}\left(g_{\sigma \mu} \dot{x}^\mu \right) - \frac{1}{2} \left( \partial_\sigma g_{\mu\nu} \right)\dot{x}^\mu \dot{x}^\nu \right).
\end{align*}But the term within the bracket is zero since $\gamma$ is a geodesic as this is the Euler-Lagrange equation for $L = \frac{1}{2} g_{\mu\nu} \dot{x}^\mu \dot{x}^\nu$.
\begin{align*}
    \frac{\dd}{\dd\lambda}\left(g_{\sigma \mu} \dot{x}^\mu \right) - \frac{1}{2} \left( \partial_\sigma g_{\mu\nu} \right)\dot{x}^\mu \dot{x}^\nu = 0\implies 0 = 2\frac{\dd}{\dd\lambda}\left(f_\mu \dot{x}^\mu \right).
\end{align*}Which is exactly the condition telling us that $f_\mu \dot{x}^\mu$ is constant.

\nocite{*}
\bibliography{bibliography_final}

\providecommand{\noopsort}[1]{}\providecommand{\singleletter}[1]{#1}%
\begin{thebibliography}{10}

\bibitem{DeRham}
Donato Bini, Christian Cherubini, Robert Jantzen, and Remo Ruffini.
\newblock De rham wave equation for tensor valued p-forms.
\newblock {\em International Journal of Modern Physics D.8.}, 12:1363–1384,
  2003.
\newblock {DOI:10.1142/S0218271803003785}.

\bibitem{Bruschi}
David~Edward Bruschi, Symeon Chatzinotas, Frank~K. Wilhelm, and Andreas~W.
  Schell.
\newblock {Spacetime effects on wavepackets of coherent light}.
\newblock 6 2021.

\bibitem{Budko}
Neil~V. Budko.
\newblock Exponential cosmological redshift in a linearly expanding universe.
\newblock 2009.

\bibitem{CarrollAstronomy}
Bradley~W. Carroll and Dale~A. Ostlie.
\newblock {\em An Introduction to Modern Astrophysics}.
\newblock Pearson Education Inc., 2013.

\bibitem{CarrollNotes}
Sean Carroll.
\newblock Lecture notes on general relativity.
\newblock
  \url{https://ned.ipac.caltech.edu/level5/March01/Carroll3/Carroll_contents.html},
  December 1997.
\newblock Accessed September 17th 2019.

\bibitem{CarrollBook}
Sean Carroll.
\newblock {\em Spacetime and Geometry: An Introduction to General Relativity}.
\newblock Pearson, first edition, 2013.

\bibitem{History}
Steven Errede.
\newblock A brief history of the development of classical electrodynamics.
\newblock
  \url{http://web.hep.uiuc.edu/home/serrede/P435/Lecture_Notes/A_Brief_History_of_Electromagnetism.pdf}.
\newblock Accessed November 8th 2019.

\bibitem{Exirifard}
Qasem Exirifard, Eric Culf, and Ebrahim Karimi.
\newblock {Towards communication in a curved space-time geometry}.
\newblock {\em {Communcations Physics}}, 4, 2021.

\bibitem{Griffiths}
David~J. Griffiths.
\newblock {\em Introduction to Electrodynamics}.
\newblock Pearson Education Inc., fourth edition, 2013.

\bibitem{Gronwald}
F.~Gronwald and J.~Nitsch.
\newblock The structure of the electromagnetic field as derived from first
  principles.
\newblock {\em IEEE Antennas and Propagation Magazine}, 43:64--79, 2001.
\newblock {10.1109/MAP.2001.951560}.

\bibitem{Hawking}
S.W. Hawking and G.F.R. Ellis.
\newblock {\em The large scale structure of space-time}.
\newblock Cambridge University Press, 1973.

\bibitem{Jackson}
John~David Jackson.
\newblock {\em Classical Electrodynamics}.
\newblock John Wiley \& Sons, Inc., third edition, 1962.

\bibitem{CurvedDipole}
B~Linet.
\newblock Dipole field solution of maxwell{\textquotesingle}s equations in the
  schwarzschild metric.
\newblock {\em Journal of Physics A: Mathematical and General}, 8(3):328--333,
  mar 1975.

\bibitem{BlackHolesNotes}
Jorma Louko.
\newblock Black holes, 2019.

\bibitem{Maldacena_1999}
Juan Maldacena.
\newblock {\em International Journal of Theoretical Physics},
  38(4):1113–1133, 1999.

\bibitem{Gravitation}
Charles~W. Misner, Kip~S. Thorne, and John~Archibald Wheeler.
\newblock {\em Gravitation}.
\newblock Princeton University Press, 2017.

\bibitem{Redshift}
Erwin Schr\"odinger.
\newblock {\em Expanding Universes}.
\newblock Cambridge University Press, Cambridge, second edition, 1956.

\bibitem{Thesis}
G.~Slezakova.
\newblock {\em Geodesic Geometry of Black Holes}.
\newblock {PhD} dissertation, The University of Waikato, Hamilton, New Zealand,
  2006.
\newblock Accessed September 17th 2019.

\bibitem{EscapeofPhotons}
J.~K. Synge.
\newblock The escape of photons from gravitationally intense stars.
\newblock {\em Monthly Notices of the Royal Astronomical Society}, 131:463,
  1965.

\bibitem{Wald}
Robert~M. Wald.
\newblock {\em General Relativity}.
\newblock The University of Chicago Press, 1984.

\bibitem{Weinberg}
Steven Weinberg.
\newblock {\em Gravitation and Cosmology: Principles and Applications of the
  General Theory of Relativity}.
\newblock John Wiley \& Sons, Inc., 1972.

\end{thebibliography}
\bibliographystyle{plain}

\end{document}